\pdfoutput=1
\documentclass[conference]{IEEEtran}

\newcommand*\circled[1]{\tikz[baseline=(char.base)]{
		\node[shape=circle,draw,inner sep=0.8pt, IndianRed1, fill=IndianRed1] (char)
		{\color{white}\scriptsize\textbf{#1}};}%
}

\PassOptionsToPackage{bookmarks=false,draft}{hyperref}

\usepackage[utf8]{inputenc}
\usepackage{textcomp}

\usepackage{xspace}
\usepackage[x11names]{xcolor}
\usepackage{graphicx}
\usepackage{pgf}
\usepackage{csquotes}

\usepackage{amsmath}
\usepackage{amssymb}
\usepackage{amsfonts}
\usepackage{mathtools}

\usepackage{booktabs}
\usepackage{array}
\usepackage{tabularx}
\usepackage{longtable}
\usepackage{adjustbox}
\usepackage{rotating}
\usepackage[flushleft]{threeparttable}
\ifCLASSOPTIONcompsoc
  \usepackage[caption=false, font=normalsize, labelfont=sf, textfont=sf]{subfig}
\else
  \usepackage[caption=false, font=footnotesize]{subfig}
\fi
\usepackage{tikz}
\usepackage{multirow}
\usepackage{threeparttable} 

\usepackage{float}
\usepackage[inline]{enumitem}
\usepackage{balance}

\usepackage{gensymb}
\usepackage{wasysym}
\usepackage{siunitx}
\usepackage[nolist]{acronym}
\usepackage[shortcuts]{extdash}

\usepackage{cite}

\usepackage{letltxmacro}

  \renewenvironment{thebibliography}[1]{%
    \begin{oldthebibliography}{#1}%
      \setlength{\parskip}{0ex}%
      \setlength{\itemsep}{0ex}%
  }%
  {%
    \end{oldthebibliography}%
  }
\definecolor{lightblue}{HTML}{2596be}

\newcommand{\etal}{\textit{et~al.}\xspace}
\newcommand{\ie}{i.e.,\xspace}
\newcommand{\eg}{e.g.,\xspace}

\newcolumntype{K}[1]{>{\raggedright\let\newline\\\arraybackslash\hspace{0pt}}m{#1}}
\newcolumntype{C}[1]{>{\centering\let\newline\\\arraybackslash\hspace{0pt}}m{#1}}
\newcolumntype{R}[1]{>{\raggedleft\let\newline\\\arraybackslash\hspace{0pt}}m{#1}}
\newcolumntype{P}[1]{>{\endgraf\vspace*{-\baselineskip}}p{#1}}

\newcolumntype{L}{>{\raggedright\arraybackslash}X}

\LetLtxMacro\oldttfamily\ttfamily
\DeclareRobustCommand{\ttfamily}{\oldttfamily\csname ttsize\endcsname}
\newcommand{\setttsize}[1]{\def\ttsize{#1}}%

\def\IEEEauthorrefmark#1{\raisebox{0pt}[0pt][0pt]{\textsuperscript{\footnotesize\ensuremath{\ifcase#1\or *\or \dagger\or \ddagger\or%
    \bullet\or \circ\or \cdot\or \times\or \checkmark\or **\or \dagger\dagger%
    \or \ddagger\ddagger \else\textsuperscript{\expandafter\romannumeral#1}\fi}}}}

\graphicspath{{./figure/}} 

\usepackage{orcidlink}
\author{
\IEEEauthorblockN{Robert Bitterling\IEEEauthorrefmark{5}\IEEEauthorrefmark{4}\orcidlink{0009-0000-4530-0670}, Christian Nettersheim\IEEEauthorrefmark{5}\IEEEauthorrefmark{4}\orcidlink{0009-0005-1504-7033}, Jörn Hees\IEEEauthorrefmark{4}\orcidlink{0000-0002-0084-8998} and Michael Rademacher\IEEEauthorrefmark{5}\IEEEauthorrefmark{4}\orcidlink{0000-0002-3721-370X}\tiny{}}
\vspace*{.28cm}
\IEEEauthorblockA{
\begin{tabular}{cc}
        {\IEEEauthorrefmark{5}Fraunhofer FKIE} &   {\IEEEauthorrefmark{4}Hochschule Bonn-Rhein-Sieg} \\
    	{Cyber Analysis \& Defense}  &  {Department of Computer Science} \\
	    {Wachtberg, Germany} & {Sankt Augustin, Germany}\\
  \small\texttt{\{firstname.lastname\}@fkie.fraunhofer.de} & \small\texttt{\{firstname.lastname\}@h-brs.de} \\
  \end{tabular}\\
  \vspace*{-.1cm}
}}

\title{A Systematic Sample Size Analysis of ML-Based Path Loss Prediction for LPWAN \\
}

\begin{document}
\bstctlcite{IEEEexample:BSTcontrol}

\setttsize{\small}
\maketitle
\begin{acronym}[]
\acro{LPWAN}{Low Power Wide Area Network}
\acro{IoT}{Internet of Things}
\acro{AI}{Artificial Intelligence}
\acro{AMSL}{Above Mean Sea Level}
\acro{LoRa}{Long Range}
\acro{NB-IoT}{Narrowband-\ac{IoT}}
\acro{LTE-M}{Long-Term Evolution for Machines}
\acro{ML}{Machine Learning}
\acro{SF}{Spreading Factor}
\acro{LoRaWAN}{Long Range Wide Area Network}
\acro{LR}{Linear Regression}
\acro{SVM}{Support Vector Machine}
\acro{ANN}{Artificial Neural Network}
\acro{RF}{Random Forest}
\acro{ADR}{Adaptive Data Rate}
\acro{SVR}{Support Vector Regression}
\acro{Bi-LSTM}{Bidirectional Long Short-Term Memory}
\acro{RBF}{Radial Basis Function}
\acro{RPP}{Received Packet Power}
\acro{LDPL}{Log-Distance Path Loss}
\acro{RMSE}{Root Mean Squared Error}
\acro{LiDAR}{Light Detection and Ranging}
\acro{MNO}{Mobile Network Operator}
\acro{k-NN}{k-Nearest Neighbors}
\acro{DBSCAN}{Density-Based Spatial Clustering of Applications with Noise}
\acro{FSPL}{Free-Space Path Loss}
\acro{LOS}{Line of Sight}
\acro{NLOS}{Non-Line-of-Sight}
\acro{LOGO}{leave-one-gateway-out}
\end{acronym}

\hyphenation{pa-ra-me-ters}
\hyphenation{environ-mental}

\begin{abstract}
Low Power Wide Area Networks like LoRa are increasingly deployed for smart city applications, requiring accurate path loss prediction for effective network planning. Traditional (empirical) propagation models often exhibit limited accuracy in these scenarios. We investigate machine learning models for LoRa path loss prediction, systematically analyzing how prediction accuracy scales with training set size using real-world measurements from an urban deployment. Our approach employs a Random Forest with LiDAR-derived terrain features and k-Nearest Neighbors with coordinate data, comparing their performance against established empirical models and specialized LPWAN models.
Under random pooled splits, both ML models consistently outperform the considered baseline models across the evaluated training-set sizes. At maximum training size, they achieve RMSE values below 6.5 dB compared to 9.7 dB for the best baseline, indicating accurate within-deployment interpolation. A leave-one-gateway-out check qualifies this result: RF shows placement-dependent transfer to held-out gateways, with moderate degradation for several gateways but larger errors for others, whereas coordinate-only k-NN degrades substantially when the gateway location is unseen.
\end{abstract}

\begin{IEEEkeywords}
\acl{LPWAN}, Propagation Modeling, Machine Learning
\end{IEEEkeywords}

\acresetall

\section{Introduction}
\label{chapter:one}
Smart city \ac{IoT} deployments require wide-area, low-power connectivity. \ac{LPWAN} technologies such as \ac{LoRa} are well suited for this setting because they provide long range, low energy consumption, and operation in unlicensed bands~\cite{Bodenhausen}.
Establishing these networks requires careful planning to meet capacity and coverage requirements~\cite{10278512}. Local authorities often have less network planning expertise than \acp{MNO}, which makes planning difficult. A key task is to estimate signal strength across locations with propagation models. Empirical models balance computational cost, required knowledge, and accuracy, but several studies report limited accuracy in \ac{LPWAN} networks \cite{Rademacher2021PathStudy,Stusek2020AccuracyScenarios,Jorke2017UrbanMHz}. This has motivated the development of refined or alternative models.
For example, Rademacher \etal conducted a large-scale measurement campaign and derived new parameters for a \ac{LDPL} model, which outperformed traditional models~\cite{Rademacher2021PathStudy}.
Despite its improved performance, this approach is not feasible for every region, as it requires a considerable amount of work and time to collect the necessary data. This constraint highlights the need for alternative approaches that can achieve high prediction accuracy with fewer data points and less domain-specific knowledge.

As an alternative, \ac{AI} and \ac{ML} methods have emerged as promising candidates for predicting signal strength in wireless networks~\cite{Seretis2021AnModeling}.
Given the cost of collecting large datasets, we ask how prediction accuracy depends on training-set size in practical smart city planning and how the models compare to traditional path-loss models.
We then examine whether these gains reflect transfer beyond the measured gateways or mainly interpolation within the existing deployment. The key contributions of this work are as follows:

\begin{itemize}[nosep]
  \item We empirically demonstrate that simple \ac{ML} models, \ac{RF} with \ac{LiDAR}-derived terrain features and coordinate-based \ac{k-NN}, outperform conventional propagation models under random pooled evaluation.
  \item We provide a systematic analysis of learning curves and training-set size effects, offering practical guidance for model selection and data collection by showing how accuracy scales with available data.
  \item We further add a \ac{LOGO} evaluation to check how performance changes when all measurements from one gateway are held out.
\end{itemize}

To facilitate credibility in our results and enable other researchers to build upon our work, we released the complete source code of this work on our GitHub page\footnote{https://github.com/mclab-hbrs/lora-bonn-ml-pathloss}.

\section{Background and Related Work}
\label{chapter:two}
In this section, we introduce empirical propagation models, discuss their challenges for \acp{LPWAN}, and then review \ac{ML} for propagation loss estimation in these networks.

Classical path loss models estimate the reduction in signal strength between a transmitter and receiver due to propagation effects. An accurate estimation of this reduction is a necessity for link budget calculations in wireless networks.
Classic empirical models (\ie Okumura–Hata~\cite{Hata1980EmpiricalServices} and COST-231 Hata~\cite{EuropeanCommission:Directorate-GeneralfortheInformationSocietyandMedia1999COSTReport}) consider influencing factors, such as the type of environment, frequency, distance, and antenna heights.
However, some of these empirical models exhibit limited accuracy in LPWAN scenarios~\cite{Stusek2020AccuracyScenarios,Jorke2017UrbanMHz,Rademacher2021PathStudy}.
Their performance can degrade outside their original calibration environments and frequency ranges.
This has motivated researchers such as Jörke \etal~\cite{Jorke2017UrbanMHz}, Stusek \etal~\cite{Stusek2020AccuracyScenarios}, and Rademacher \etal~\cite{Rademacher2021PathStudy} to develop or refine empirical models specifically based on particular measurement campaigns or deployment scenarios. The process of creating or refining these models typically follows the same pattern: identify a shortcoming with current models, conduct a measurement campaign, preprocess the data, derive or refine a model, and evaluate it against the measurements.

The described empirical modeling process closely resembles supervised \ac{ML} workflows. In these workflows, models are trained on labeled data to predict quantities such as path loss~\cite{Seretis2021AnModeling}. The key difference is that, while empirical models typically use a fixed functional form and adjust only certain parameters, ML algorithms can learn more flexible relationships from data without a predefined structure. Recent surveys, such as the one by Seretis and Sarris~\cite{Seretis2021AnModeling}, have shown that ML models, including both \ac{ANN}-based and non-\ac{ANN} models like \ac{RF} and \ac{k-NN}, can achieve higher accuracy than traditional empirical models for modeling radio wave propagation. \acl{RF}s are frequently highlighted for their strong performance.

Although there are numerous studies applying \ac{AI} techniques in wireless communications, a review of the literature reveals that only a limited number have focused specifically on \ac{ML} to model radio wave propagation in \ac{LoRa} scenarios. Some of the identified studies focus on \ac{ML} and \ac{LoRa}~\cite{Ballestrin2024ExploringNetworks,Liu2021DeepLoRa:LPWAN}, while others use \ac{ML} only as a support tool rather than as their primary research focus~\cite{Alobaidy2022Low-Altitude-Platform-BasedEnvironment,Gonzalez-Palacio2023Machine-Learning-BasedEnhancement}. 

Alobaidy \etal~\cite{Alobaidy2022Low-Altitude-Platform-BasedEnvironment} developed a hybrid ML model for water quality monitoring in a tropical environment. The approach combined free space path loss with linear regression and a bagged trees ensemble, using 17 radio and environmental features. The data were preprocessed with standard ML methods and augmented with ray tracing and a 3D building map. Their model outperformed conventional models.

Gonzalez-Palacio \etal~\cite{Gonzalez-Palacio2023Machine-Learning-BasedEnhancement} aimed to save energy in \ac{LoRa} networks by proposing an \ac{ADR} algorithm that incorporates environmental factors. They evaluated multiple \ac{ML} models, including multiple linear regression, \acp{ANN}, support vector regression, and \acp{RF}, to predict path loss statistics. \acp{RF} performed best.

Ballestrin \etal~\cite{Ballestrin2024ExploringNetworks} developed and benchmarked \ac{ML} models for \ac{LoRa} path loss prediction using \acp{ANN}, \acp{RF}, \ac{SVR}, \ac{LR}, XGBoost, and decision trees. The models used a dataset enriched with elevation data and were compared to Okumura and log-distance baselines. All \ac{ML} models except \ac{SVR} with a sigmoid kernel outperformed the traditional models, with \acp{RF}, XGBoost, and decision trees performing best. The study reported limited details on preprocessing and feature enrichment.

Liu \etal~\cite{Liu2021DeepLoRa:LPWAN} addressed the limits of dominant land cover features by developing a deep neural network based on \ac{Bi-LSTM}. They collected 30,000 urban data points using two gateways and six end nodes. The model used sequences of classified land cover types, distance, and height difference as inputs. Land cover was automatically classified from remote sensing images with \ac{SVM} using a \ac{RBF} kernel, distinguishing \ac{LOS} and \ac{NLOS} types such as trees, buildings, water, and roads. Training on these features allowed the model to approximate propagation behavior and generalize to new environments. It halved the error of conventional methods.

In contrast to previous studies, our work focuses on the amount of training data required for accurate ML-based LoRa propagation-loss prediction. Rather than constructing complex or hybrid ML architectures or employing deep learning models, we emphasize simple, easy-to-use ML models. Furthermore, our study does not concentrate solely on maximizing prediction performance or benchmarking different ML approaches. Instead, we investigate these models as an extension to Rademacher \etal's~\cite{Rademacher2021PathStudy} work. Specifically, we examine how training-set size affects model performance.

\section{Methodology}
\label{chapter:three}

\begin{figure*}[ht]
	\centering
	\centering
	\includegraphics[width=\linewidth]{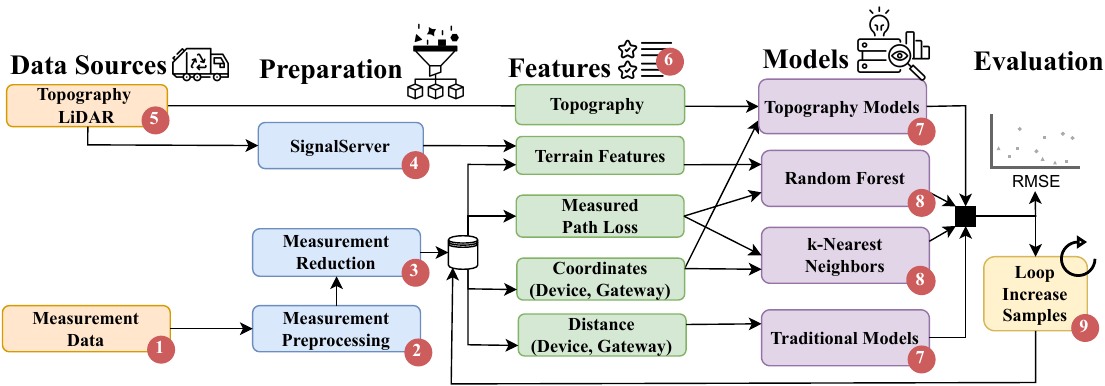}
	\caption{Overview of the experimental pipeline. LiDAR and measurement data are prepared. From these sources, we derive features (converted LiDAR/topographic inputs, SignalServer outputs, measured path loss, coordinates, and distances) to train and evaluate models (deterministic topography-based, traditional empirical, RF, and k-NN). Performance is assessed by RMSE. The experiment is repeated across multiple training set sizes.}
	\label{fig:experiment}
\end{figure*}

To examine how training-set size affects LoRa path-loss prediction, we designed an experimental setup using real-world measurements and terrain data. Our approach emphasizes a comparative evaluation of traditional and ML-based models. We detail dataset preparation, features, model selection, and experimental setup in the following. An overview is shown in~\autoref{fig:experiment}, and we refer to the numbered steps.

We use the final post-processed dataset from the study by Rademacher \etal~\cite{Rademacher2021PathStudy} rather than collecting new measurements~\circled{1}. The data were collected in Bonn, Germany, an urban area of approximately 330,000 inhabitants.
Uplink \ac{RPP} was recorded from outdoor end devices mounted on garbage trucks over 230 days, using nine gateways citywide at \SI{868}{\mega\hertz}, \SI{125}{\kilo\hertz} bandwidth, and \ac{SF} 12. After the original cleaning, 131,057 samples remained. For this work, we removed samples and gateways outside the city limits, reducing the set to 114,465 samples and eight gateways~\circled{2}. From measured \ac{RPP} and known system parameters (\eg antenna gains), we computed path loss. To reduce redundancy, we removed duplicates and applied \ac{DBSCAN}~\cite{ester1996density} to sender and receiver coordinates, grouping spatially proximate samples and consolidating overlaps, yielding 64,294 samples~\circled{3}.

\begin{figure}[hb]
    \centering
    \includegraphics[width=3.4in]{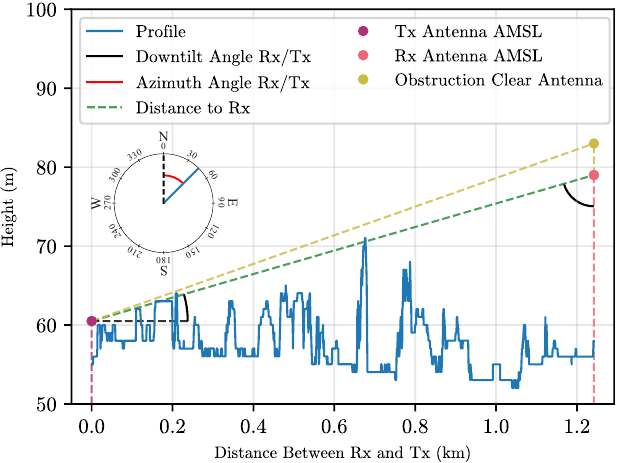}
    \caption{Illustration of selected terrain features used as input variables for the \ac{RF} models. Displayed features are: \enquote{Tx antenna height \ac{AMSL}}, \enquote{Rx antenna height \ac{AMSL}}, \enquote{Downtilt angle to Tx}, \enquote{Downtilt angle to Rx}, \enquote{Distance to Rx}, \enquote{Rx antenna height to clear all obstructions detected}, \enquote{Azimuth to Tx}, and \enquote{Azimuth to Rx}. The \enquote{Profile} curve is included for illustration only and is not used as an input feature. All features were generated using \enquote{SignalServer}. Additional model features not displayed include: \enquote{Rx antenna height to clear the first Fresnel zone}, \enquote{Rx antenna height to clear 60\% of the first Fresnel zone}, and \enquote{FSPL}.}
    \label{fig:features_rf}
\end{figure}

A key idea in this work is to use terrain-based features as input to \ac{ML}-based \ac{LPWAN} modeling. To obtain terrain-based features, we employ~\circled{4} \enquote{SignalServer}\footnote{https://github.com/lmux/Signal-Server/tree/85589ed6}, a multithreaded RF-propagation software based on \enquote{SPLAT!}\footnote{https://www.qsl.net/kd2bd/splat.html}. \enquote{SignalServer} can produce 360° polar coverage maps as well as 2-D point-to-point path profiles. LiDAR topographic data (provided by the state of NRW in Germany) were converted from LAZ (EPSG:25832) to ASCII grid (EPSG:4326) for \enquote{SignalServer}~\circled{5}. From the path profiles for all data points, we derive features including antenna height \ac{AMSL} at transmitter and receiver, receiver antenna height required to clear the first Fresnel zone, receiver antenna height required to clear all detected obstructions, receiver antenna height required to clear 60\% of the first Fresnel zone, \ac{FSPL}, downtilt angles to transmitter and receiver, azimuths to transmitter and receiver, and distance to receiver (\autoref{fig:features_rf}). After preprocessing and feature engineering \circled{6}, the available inputs are topographic data, terrain-derived features, coordinates, and distance, with path loss as the target variable.

Traditional empirical and terrain-based models serve as baseline approaches in our experiments~\circled{7} and include both widely used empirical models such as Okumura-Hata, ECC33~\cite{ITU-R_P529-3ECC33}, ITM~\cite{Hufford1982ITM}, and ITWOM~\cite{Shumate2010ITWOM}, and specialized models such as Dortmund~\cite{Jorke2017UrbanMHz}, Ghent~\cite{CallebautPerre2020}, Oulu~\cite{Petajajarvi2016OnTechnology}, and Bonn~\cite{Rademacher2021PathStudy}. Following the protocol of Rademacher \etal~\cite{Rademacher2021PathStudy}, we selected the same set of models to ensure consistency and comparability between studies. This selection strategy reflects two common scenarios encountered in radio propagation research: the use of established, widely recognized models, and the development or adoption of site-specific models based on dedicated measurement campaigns. Models such as ECC33, COST-Hata, Ericsson, Egli, ITM, and ITWOM were computed using \enquote{SignalServer}, while Okumura-Hata and specialized models (Dortmund, Ghent, Oulu, and Bonn) were implemented directly by us.

As ML models, we selected \ac{k-NN} and \ac{RF} for methodological and practical reasons~\circled{8}. Prior work suggests that, outside deep learning, common regressors yield similar path-loss accuracy. We chose \ac{k-NN} for simplicity and interpretability and \ac{RF} for strong prior performance~\cite{Ballestrin2024ExploringNetworks,Gonzalez-Palacio2023Machine-Learning-BasedEnhancement} and out-of-the-box usability. Both have low computational overhead.

For k-NN, we use only the latitude and longitude of transmitter and receiver as features, with path loss as the target. This defines a simple, low-input ML baseline. The coordinates implicitly encode link distance and local spatial context, while avoiding the terrain-derived features used by RF. This allows us to compare a coordinate-based learner with both traditional low-input propagation models and the terrain-feature-based RF model. The value k = 31 with distance weighting was selected via grid search over $k \in$ \{1, 3, 5, 7, 9, 11, 15, 21, 31, 41, 51, 61, 71, 81, 91, 101, 151, 201, 251, 301, 351\}, testing uniform and distance weighting. Distance weighting with $k=31$ yielded the lowest validation RMSE. However, to address potential oversmoothing and poor performance at low sample sizes, a variant with $k=7$ was also evaluated in iterations with limited training data. Although incorporating the explicit distance between transmitter and receiver as a feature in \ac{k-NN} was considered, we opted not to do so, as the coordinate-based approach essentially provides a spatial lookup and maintains simplicity.

To go beyond simple region-bound coordinate features and to provide inputs comparable to terrain-based models such as ITM and ITWOM, we used the terrain information available through SignalServer. ITM and ITWOM require topography data as input, which SignalServer processes directly to compute their propagation predictions. The same SignalServer path-profile analysis that enables these traditional models also provides the terrain-based features described above, including elevation data, Fresnel zone clearances, and azimuth information. To leverage these engineered features in a machine learning context, we employed \ac{RF} (using YDF\footnote{https://ydf.readthedocs.io/en/stable/}), which learns patterns from these terrain-derived features rather than processing the raw topographic data.

Our primary evaluation measures how accuracy scales with training-set size. Following~\cite{Ballestrin2024ExploringNetworks}, we used a 75/25 random pooled split: a test set with a fixed size of 16,074 samples was held out, and training subsets of increasing size were drawn at random from the remaining data~\circled{9}, with each configuration repeated by resampling the training subset. This pooled split cannot indicate how far the models transfer beyond the measured gateways. Rather than relying on a second, independent dataset, we use a \ac{LOGO} check at the largest training size to gain initial insight into this: all samples from one gateway are held out for testing and the model is trained on the remaining gateways. We report each \ac{LOGO} metric as the unweighted mean over the eight held-out gateways, so that every gateway contributes equally rather than in proportion to its number of measurements.

\section{Evaluation Results}
\label{chapter:four}
\begin{table}[!t]
	\centering
	\begin{threeparttable}
		\footnotesize
		\caption{RMSE (\si{\deci\bel}) comparison of models across training-set sizes.}
		\label{tab:rmse}
		\begin{tabular}{@{}c@{\hspace{8pt}}ccc@{\hspace{12pt}}cccc@{}}
			\toprule
			\textbf{Sample} & \multicolumn{3}{c@{\hspace{12pt}}}{\textbf{This Work}} & \multicolumn{4}{c@{}}{\textbf{Baseline Models}} \\
			\cmidrule(lr){2-4} \cmidrule(l){5-8}
			\textbf{Size} & \textbf{RF} & \textbf{k-7} & \textbf{k-31} & \textbf{Bonn} & \textbf{C. Ha.} & \textbf{ITM} & \textbf{ITW.} \\
			\midrule
			100     & \textbf{8.36} & 9.56 & 10.02 & \multirow{6}{*}{9.68} & \multirow{6}{*}{12.39} & \multirow{6}{*}{21.75} & \multirow{6}{*}{22.58} \\
			200     & \textbf{8.07} & 8.83 & 9.37  &                       &                       &                       &  \\
			500     & \textbf{7.74} & 8.11 & 8.47  &                       &                       &                       &  \\
			4,500   & 7.02 & \textbf{6.95} & \textbf{6.95}  &                       &                       &                       & \\
			15,000  & 6.65 & 6.58 & \textbf{6.47}  &                       &                       &                       &  \\
			48,220  & 6.34 & 6.32 & \textbf{6.15}  &                       &                       &                       &  \\
			\bottomrule
		\end{tabular}
		\begin{tablenotes}
			\footnotesize
			\item[] Note: The test-set size was held constant for all experiments. Baseline models (Bonn, COST-Hata, ITM, ITWOM) were not trained on the dataset. Therefore, their RMSE values remain approximately constant across training-set sizes. Minor differences arise from random sampling, not systematic changes in model effectiveness.
		\end{tablenotes}
	\end{threeparttable}
\end{table}

We evaluated a range of models, including \ac{RF} and \ac{k-NN} ($k=7$ and $k=31$), as described in Section~\ref{chapter:three}. To maintain clarity in the results section, we report results for a representative subset of the evaluated models.

\autoref{tab:rmse} presents the \ac{RMSE} of various models, comparing traditional models with \acl{ML} models for training-set sizes ranging from 100 to 48,220. The models discussed include the COST 231 Hata model, the LDPL Bonn model, and the terrain-based ITM and ITWOM models, together with the ML models \ac{k-NN} ($k=7$ and $k=31$) and the \ac{RF} model.
At small sample sizes, the LDPL Bonn model achieves an RMSE comparable to the k-NN models, while the RF model outperforms all traditional models. As the sample size increases, the ML models, particularly RF and k-NN with $k=31$, show substantial improvements in RMSE and reliably outperform the baseline models. The baseline models’ RMSE values remain nearly constant, with Bonn around \SI{9.68}{\deci\bel}, COST 231 Hata at \SI{12.39}{\deci\bel}, ITM at \SI{21.75}{\deci\bel}, and ITWOM at \SI{22.58}{\deci\bel}. In contrast, the RF and k-NN models reduce their RMSE values to below \SI{6.5}{\deci\bel} at the largest sample sizes.
Overall, under the random pooled evaluation, the \ac{ML} models consistently outperform the considered traditional models across the evaluated training-set sizes.
\begin{figure}[!t]
    \centering
    \includegraphics[width=3.4in]{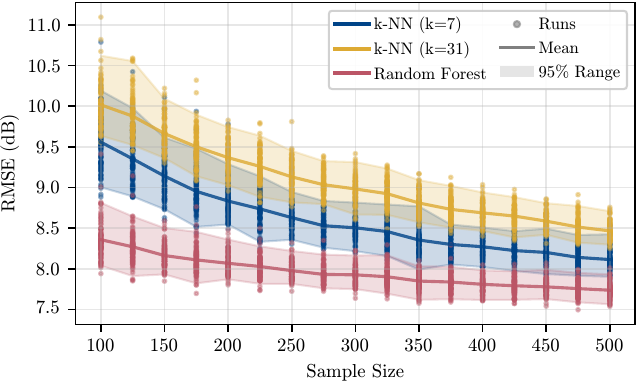}\\
    \includegraphics[width=3.4in]{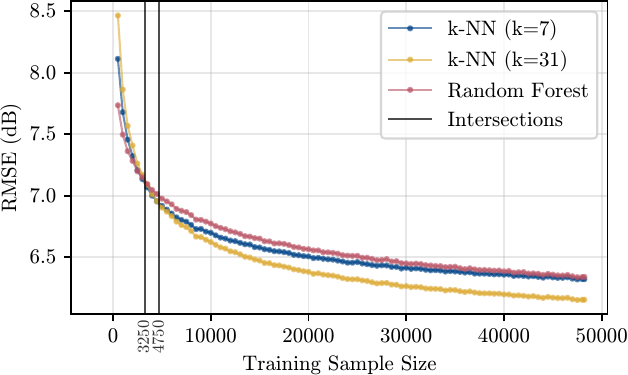}
    \caption{Mean \ac{RMSE} (\si{\deci\bel}) versus training-set size. Top: sizes 100--500 over 100 independent runs, with mean lines and 95\% envelopes; variability shrinks as the sample size grows and \ac{RF} leads across this range. Bottom: sizes 500--48{,}220, where accuracy improves for all models with diminishing returns; both k-NN variants overtake \ac{RF} at approximately 3{,}250 samples, and $k=31$ overtakes $k=7$ at approximately 4{,}750 samples.}
    \label{fig:rmse_line_plot_50_500_max}
\end{figure}

\autoref{fig:rmse_line_plot_50_500_max} provides a more detailed analysis of machine learning models. For training-set sizes between 100 and 500, the RF model consistently outperforms both k-NN models. The performance gap between the RF and k-NN models narrows as the sample size increases. Both k-NN models begin with similar RMSE values, but the model with $k=7$ improves faster than $k=31$. This is likely due to $k=7$ being more appropriate for smaller sample sizes. The improvement is substantial, with the k-NN models achieving a reduction in RMSE of approximately \SI{1.5}{\deci\bel} and the RF model improving by about \SI{0.6}{\deci\bel} as the sample size increases to 500. Although the central tendency of the RMSE curves is favorable, the spread is non-negligible, especially for k-NN: the k-NN bands overlap substantially at small sample sizes, whereas the RF band intersects those of the k-NN models only rarely and primarily at larger sample sizes. Overall, the bands narrow with increasing sample size, indicating reduced variability.

When shifting from a detailed view to coarser increments, the trend of decreasing mean RMSE persists. Up to roughly 3,250 samples the RF maintains the lowest RMSE. At this point both k-NN models overtake the RF. The first vertical reference line marks this intersection. At around 4,750 samples the k-NN with $k=31$ overtakes the other k-NN, marked by the second vertical line. From this stage onward, the k-NN with $k=7$ and the RF model exhibit closely matched RMSE values, reaching \SI{6.32}{\deci\bel} and \SI{6.34}{\deci\bel}, respectively. At the maximum training-set size, the k-NN model with $k=31$ achieves the lowest RMSE at \SI{6.15}{\deci\bel}. Overall, a clear improvement trend emerges with diminishing returns: for RF, RMSE improves by \SI{0.62}{\deci\bel} from 100 to 500 samples, by \SI{1.08}{\deci\bel} from 500 to 15,000 samples, and by only \SI{0.31}{\deci\bel} from 15,000 to 48,220 samples.

\begin{figure}[!t]
    \centering
    \includegraphics[width=3.4in]{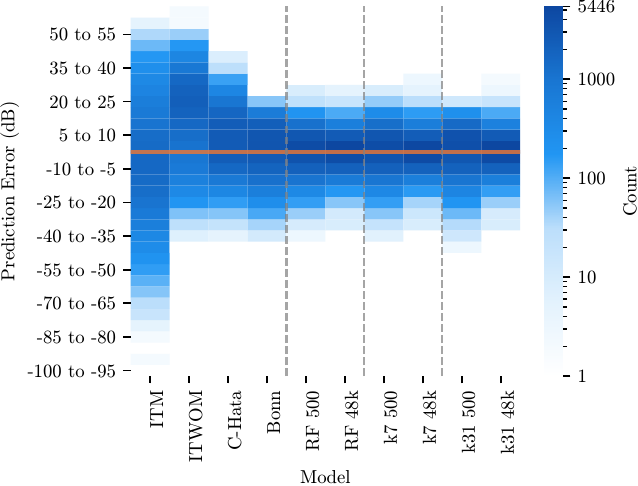}
    \caption{Distribution heatmap of path-loss prediction errors (prediction error, \si{\deci\bel}) from one iteration across prediction error bins (\SI{5}{\deci\bel} per bin) for multiple models. Logarithmic color scale (white = zero occurrences; darker blue = higher frequency). Abbreviations: ITM = Irregular Terrain Model; ITWOM = Irregular Terrain with Obstructions Model; C-Hata = COST-Hata; RF = Random Forest; k7/k31 = k-NN ($k=7$ and $k=31$). Parentheses indicate training set sizes (500 vs.\ 48k).}
    \label{fig:residual_distribution_heatmap}
\end{figure}

RMSE is an aggregate measure and does not reveal the distribution of individual errors, which \autoref{fig:residual_distribution_heatmap} addresses. In general, apart from the direction of underestimation or overestimation, the prediction error behavior of the models aligns with the trends observed in the performance line plots. Among baseline models, most tend to underestimate path loss, except ITM, whose distribution is centered near zero but has a wider spread, especially on the overestimation side. ITWOM is less dispersed but shifted toward positive prediction errors, indicating consistent underestimation. Non-terrain-based baselines show smaller overall spreads than terrain-based models.
For \ac{ML} models (\ac{RF}, k-NN with k = 7 and k = 31), we show training sizes of 500 and 48,220. Their error distributions are more compact, with narrower and darker central bins, than the baselines, although at 500 samples the errors of all model distributions are a bit more spread out. At the maximum sample size, the maximum spread of the prediction errors remains relatively unchanged, but the central error distribution narrows. Large, infrequent prediction errors persist, indicating either inherent variability in the measurements or remaining unmodeled factors. The Bonn model, a domain-specific model derived from a variant of the dataset~\cite{Rademacher2021PathStudy}, stands out among the baseline models, as its prediction error distribution closely matches that of the \ac{ML} models trained with 500 samples in both spread and central tendency.

\begin{table}[!t]
	\centering
	\begin{threeparttable}
		\footnotesize
		\caption{RMSE (\si{\deci\bel}) for the random pooled split versus \acl{LOGO} (\ac{LOGO}) at the largest training size.}
		\label{tab:logo}
		\begin{tabular*}{\columnwidth}{@{\extracolsep{\fill}}lccc@{}}
			\toprule
			\textbf{Model} & \textbf{Random} & \textbf{\ac{LOGO}} & \textbf{$\Delta$} \\
			\midrule
			RF            & 6.34 & 8.92  & +2.58 \\
			k-NN ($k=7$)  & 6.32 & 13.73 & +7.41 \\
			k-NN ($k=31$) & 6.15 & 13.28 & +7.13 \\
			\bottomrule
		\end{tabular*}
		\begin{tablenotes}
			\footnotesize
			\item[] Note: \ac{LOGO} trains on 7 of 8 gateways and tests on the held-out one, averaged with equal weight across all 8 gateways. The random baseline is the largest-size (48,220-sample) result from \autoref{tab:rmse}.
		\end{tablenotes}
	\end{threeparttable}
\end{table}

To gain a first insight into how far the models transfer beyond the measured gateways, we perform a \ac{LOGO} evaluation. \autoref{tab:logo} compares the random split and \ac{LOGO} at the largest training size. Both k-NN variants degrade sharply, roughly doubling their RMSE from about \SI{6.3}{\deci\bel} to about \SI{13.5}{\deci\bel}. This is expected, as the coordinate-only \ac{k-NN} acts as a spatial lookup: no training samples share the held-out gateway coordinate, leaving \ac{k-NN} without nearby receiver-side neighbors for interpolation. This supports the interpretation that coordinate-only k-NN mainly captures spatial interpolation within the measured deployment rather than transferable propagation behavior. The \ac{RF} model degrades far less, from \SI{6.34}{\deci\bel} to \SI{8.92}{\deci\bel} (an increase of \SI{2.58}{\deci\bel}), indicating that its terrain-derived features transfer partially to unseen gateways.

\begin{figure}[!t]
    \centering
    \includegraphics[width=3.4in]{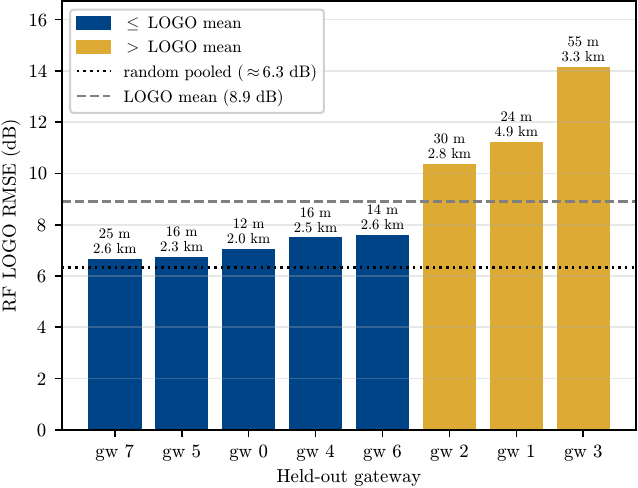}
    \caption{RF path-loss RMSE for each held-out gateway under \ac{LOGO} (trained on the other seven gateways), sorted by error and annotated with the gateway's antenna height and mean link distance. The dotted line marks the random pooled error (about \SI{6.3}{\deci\bel}), the dashed line the eight-gateway \ac{LOGO} mean. Five held-out gateways remain close to the random pooled error, whereas three gateways with larger antenna height or mean link distance drive the average up.}
    \label{fig:logo_per_gateway}
\end{figure}

This aggregate, however, hides substantial gateway-to-gateway variation. \autoref{fig:logo_per_gateway} shows the RF error for each held-out gateway. For five of the eight gateways, the RMSE remains between \SI{6.6}{\deci\bel} and \SI{7.6}{\deci\bel}, close to the random pooled error of about \SI{6.3}{\deci\bel}. This suggests that held-out gateways with characteristics similar to the training gateways can be predicted with only moderate degradation relative to the random pooled result. The remaining three gateways raise the average considerably, reaching \SIlist{10.4;11.2;14.2}{\deci\bel}. These three gateways stand out by their site characteristics: an unusually large antenna height of \SI{55}{\meter}, a mean link distance of \SI{4.9}{\kilo\meter}, and an antenna height of \SI{30}{\meter}.

\section{Discussion and Future Work}
\label{chapter:five}

Across random pooled splits \ac{ML} models outperform the considered empirical and terrain-based baselines. \ac{RF} is strongest at very small sample sizes and remains favorable up to roughly 3,250 samples. With more data, coordinate-based \ac{k-NN} becomes competitive and achieves the lowest \ac{RMSE} at the largest training size, albeit by a small margin. This behavior is expected because \ac{k-NN} acts mainly as a spatial interpolation method: as sampling density increases, the nearest measured links tend to lie closer to each prediction location. The gain per additional sample diminishes, however, indicating a practical limit to the accuracy attainable from more data alone.

The pooled-split results, however, can overestimate performance when the target gateway was not represented during training, since links from the same gateways can appear in both training and test data. Our LOGO check, which isolates this transfer setting, shows a more nuanced result. RF remains competitive on average and transfers with moderate degradation to held-out gateways that are similar to those in the training data. For some gateways, errors increase significantly. In our data, these high-error cases coincide with unusual antenna heights or different link-distance distributions. Since we did not perform a dedicated causal analysis or pre-deployment outlier-detection study, we treat these characteristics as descriptive indicators rather than confirmed causes.

For practitioners, model choice should reflect the available data, the target accuracy, and the effort of feature extraction. With only a few hundred labeled links, \ac{RF} yields the best accuracy, at the cost of deriving terrain features. With a few thousand samples or more, coordinate-only \ac{k-NN} becomes equally competitive for within-deployment interpolation while requiring only link coordinates. Beyond roughly 3,250 samples the two lie within about \SI{0.2}{\deci\bel}, a gap that should be interpreted with caution given the limited hyperparameter tuning and the differing feature sets. When data or compute is highly constrained, a traditional empirical model such as COST-231 Hata, which requires no local measurements, may still suffice.

These findings come with limitations. They are based on one urban region, environment, technology, and frequency, so whether terrain-feature models transfer across cities, technologies, or frequencies remains open. We also did not perform extensive hyperparameter tuning, which could shift absolute errors and the small differences between \ac{ML} models.

Two directions follow for future work. First, for within-deployment interpolation, the central question is how to design a small measurement campaign that still achieves the maximum tolerable \ac{RMSE}, selecting sampling locations that best represent the environment while reducing the variability seen at low sample sizes. Second, replication on independent deployments in other cities, frequencies, and technologies would test the external validity left open above.

\section{Conclusion}
We evaluated simple machine learning models for LPWAN path loss prediction using real-world LoRa measurements and LiDAR-derived terrain features in an urban setting.
Under random pooled evaluation, both models consistently outperformed the traditional empirical and terrain-based baselines across training-set sizes, reaching \ac{RMSE} below \SI{6.5}{\deci\bel} versus about \SI{9.7}{\deci\bel} for the best baseline, with clear diminishing returns as data grew.

These results show that data-driven models deliver accurate path-loss estimates for within-deployment interpolation with modest data requirements. The LOGO check qualifies this finding: RF shows partial, placement-dependent transfer to newly added gateways, whereas coordinate-only k-NN largely reflects interpolation within the measured deployment.

\bibliographystyle{IEEEtran}

\bibliography{references.bib}

\end{document}